\documentclass[aip,graphicx]{revtex4-1}
\usepackage[T1]{fontenc}
\usepackage[latin1]{inputenc}
\usepackage{graphicx}
\usepackage{setspace}
\usepackage{color}
 
\makeatletter

\newcommand{\TAE}{TAE}
\newcommand{\TAEs}{TAEs}
\newcommand{\MHD}{MHD}
\newcommand{\NOVA}{NOVA}
\newcommand{\DGAE}{DGAE}
\newcommand{\CKA}{CKA}

\makeatother
\begin{document}

\title
{Comparison of methods for numerical calculation of continuum damping}

\author{G.W. Bowden}
\affiliation{Research School of Physical Sciences and Engineering, Australian National University, Acton 0200, ACT Australia}
\author{A.~K{\"{o}}nies}
\affiliation{Max-Planck-Institut f\"{u}r Plasmaphysik, EURATOM-Association, D-17491 Greifswald, Germany}
\author{M.J.~Hole}
\affiliation{Research School of Physical Sciences and Engineering, Australian National University, Acton 0200, ACT Australia}
\author{N.N.~Gorelenkov}
\affiliation{Princeton Plasma Physics Laboratory, P.O. Box 451, Princeton, New Jersey 08543-0451, USA}
\author{G.R. Dennis}
\affiliation{Research School of Physical Sciences and Engineering, Australian National University, Acton 0200, ACT Australia}

\begin{abstract}
Continuum resonance damping is an important factor in determining the stability of certain global modes in fusion plasmas. A number of analytic and numerical approaches have been developed to compute this damping, particularly in the case of the toroidicity-induced shear Alfv\'{e}n eigenmode. This paper compares results obtained using an analytical perturbative approach with those found using resistive and complex contour numerical approaches. It is found that the perturbative method does not provide accurate agreement with reliable numerical methods for the range of parameters examined. This discrepancy exists even in the limit where damping approaches zero. When the perturbative technique is implemented using a standard finite element method, the damping estimate fails to converge with radial grid resolution. The finite elements used cannot accurately represent the eigenmode in the region of the continuum resonance, regardless of the number of radial grid points used. 

\end{abstract}


\maketitle

\section{Introduction}

Lightly damped global modes can  exist in magnetically confined plasmas due to various couplings between different poloidal and toroidal harmonics \cite{excitation_of_global_eigenmodes_of_the_Alfven_wave_in_tokamaks}. These couplings result from asymmetry in the plasma geometry and create an effective potential well for the modes, analogous to defects in optics. Such global modes include the toroidicity-induced Alfv\'{e}n eigenmode (\TAE), which is due to the coupling between different poloidal harmonics resulting from toroidicity \cite{low_n_shear_Alfven_spectra}. Resonances occur where the frequency of such a global solution to the magneto-hydrodynamic (\MHD) equations coincides with that of the local continuum solution. These resonances result in transfer of energy from the global mode to the heavily damped continuum modes. Continuum resonance damping can be one of the main sources of damping for these global modes. As such, this damping plays an important role in determining whether global modes will be driven unstable due to resonances with fast particles \cite{excitation_of_TAEs_by_fusion_alphas}. Damping due to interaction with the continuum can result in transfer of energy from the \TAE\ to heat narrow layers of plasma surrounding resonances \cite{theory_of_plasma_heating_by_low_frequency_waves}. Continuum resonance damping at edge resonances typically inhibits the coupling of energy from outside the plasma to its core, making it difficult to use \TAEs\ to efficiently couple energy to the core plasma.

Continuum resonance damping has been studied using both analytic and numerical techniques. In ideal \MHD\ singularities exist at continuum resonances and their proper treatment is determined by the causality condition, similar to the analysis of Landau damping \cite{landau_damping}. For high toroidal mode number, $n$, and large aspect ratio, $A$, continuum resonance damping has been calculated analytically using asymptotic matching \cite{continuum_damping_of_high_n_TAWs}. Under these same assumptions, continuum damping has also been analysed by applying a perturbative representation of the resonance following a ballooning transformation of the coordinates \cite{continuum_damping_of_ideal_TAEs,resonant_damping_of_TAEs_in_tokamaks}. Alternatively, for low $n$, continuum damping can be calculated based on the perturbation of a quadratic form constructed from the wave equation \cite{continuum_damping_of_low_n_TAEs}. A commonly used numerical approach is to calculate the imaginary part of the mode frequency in the limit as plasma resistivity is reduced to zero \cite{damping_of_GAEs_in_tokamaks_due_to_resonant_absorption}. Finally, solving the equations for global modes over a suitable contour in the complex plane also gives damping as an imaginary component of the mode frequency \cite{computational_approach_to_continuum_damping_in_3D_published}. This method is less numerically intensive than a resistive calculation, though requires analytic continuations for equilibrium quantities. This paper compares the predictions of the perturbative technique for cases of low $n$ \TAEs\ with those of complex contour and resistive techniques.

Analytic approximations of continuum damping are of interest because they can illuminate the dependence of damping on equilibrium parameters. However, little information exists regarding how these theories compare with numerical results. Section \ref{sec:perturbative} of this paper summarises the perturbative technique developed by Berk \textit{et al.}\ \cite{continuum_damping_of_low_n_TAEs}, the complex contour technique developed by K\"{o}nies and Kleiber \cite{computational_approach_to_continuum_damping_in_3D_published} and the resistive technique. Subsequently, section \ref{sec:perturbative_vs_contour} compares the damping estimates obtained using this analytic technique for simplified geometry to those found using a numerical technique solving the global mode equations over a complex contour. In section \ref{sec:perturbative_vs_resistive}, the results from two \MHD\ codes which respectively implement perturbative and resistive methods are compared for a more complex geometry.

\section{Overview of methods for calculating continuum resonance damping} \label{sec:perturbative}

\subsection{Perturbative method}

A perturbative technique for calculating the damping of a \TAE\ due to continuum resonances was developed by Berk \textit{et al.}\ \cite{continuum_damping_of_low_n_TAEs}. Their technique is based on a simplified coupled mode equation for \TAEs\ which is derived for a large aspect ratio circular tokamak. That equation can be expressed as follows:
\begin{eqnarray} 
\noindent \frac{d}{dr}\left [r^3\left (\frac{\omega^2}{{v_A}^2}-k_{\parallel m}^2 \right ) \frac{dE_m}{dr}\right]+\frac{d}{dr}\left (\frac{\omega^2}{{v_A}^2} \right )r^2E_m-\left (m^2-1 \right )\left(\frac{\omega^2}{v_A^2}-k_{\parallel m}^2 \right )rE_m && \nonumber \\
\noindent +\frac{d}{dr}\left [\frac{5\epsilon r^4}{2a}\epsilon \left (\frac{dE_{m+1}}{dr}+\frac{dE_{m-1}}{dr} \right ) \right ]=0 \label{eq:wave_equation}
\end{eqnarray}
where $E_m$ is the $m$'th poloidal Fourier component of the quantity $\frac{\delta \Phi}{r}$ and $\delta \Phi$ is the perturbation of the electric potential associated with the shear Alfv\'{e}n wave. Note that the gauge is set such that the magnetic vector potential $\mathbf{A}$ is perpendicular to the magnetic field. The radial and poloidal angle coordinates in the flux-type straight-field-line coordinates defined by Berk \textit{et al.} are $r$ and $\theta$ respectively \cite{continuum_damping_of_low_n_TAEs}. The inverse aspect ratio is is $\epsilon=\frac{a}{R_0}$ where $a$ is the minor radius of a tokamak and $R_0$ is its major radius.

A ``flux'' function $C_m$ is defined as the sum of the terms in square brackets in equation (\ref{eq:wave_equation}) \cite{continuum_damping_of_low_n_TAEs}. This second order differential equation is decomposed into a pair of coupled differential equations in which the derivative of each of the two variables $E_m$ and $C_m$ can be expressed as a linear function of the other. At resonances, the definition for $C_m$ becomes non-invertible, though a Frobenius expansion reveals that $C_m$ remains finite at this point. Hence, the value of $C_m$ at the resonance can be used to calculate the discontinuity in $E_m$ at that point.

The original wave equation is also used by Berk \textit{et al.} to derive an expression relating a quadratic form in terms of $E_m$ and $C_m$ to discontinuities in $E_m$ at resonant singular points. Using Einstein notation,
\begin{equation}
G \left (\omega,E_m \right ) = \sum_{j} \left [ {\lim_{\delta \to 0+} \left (  E_m \left ( r - \delta \right ) - E_m \left ( r + \delta \right ) \right ) C_m\left ( r_{s,j} \right ) } \right ]
\end{equation}
for
\begin{equation}
G \left (\omega,E_m \right ) \equiv \mathcal{P} \int_{0}^{a} \left [ \frac{d }{d r}{\left (E_m C_m \right )} \right ] dr
\end{equation}
In this equation $\mathcal{P}$ denotes the Cauchy principal value of the integral with respect to the singular points. The locations of the singularities are $r_{s,j}$. Assuming that damping is a small first order correction to the eigenfrequency $\omega$, this quantity and the eigenfunction, $E_m$ are respectively taken to be the sum of an unperturbed part ($E_m^{ \left ( 0 \right ) }$ and $\omega^{\left (0 \right )}$) and a perturbation resulting from interaction with the continuum at the resonances ($\delta E_m$ and $\delta \omega$). In this formalism, $E_m^{ \left ( 0 \right ) }$ and $\omega^{\left (0 \right )}$ are solutions to the \TAE\ equation assuming that the wave function is continuous across singularities due to resonances. In contrast, $\delta E_m$ is allowed to be discontinuous across resonances, with a step change in its imaginary component occurring at these points. These discontinuities can be calculated based on analytic continuation of $E_m^{ \left (0 \right ) }$ in accordance with the causality condition. Variation in $G$ with $\omega$ can then be equated to an expression involving these discontinuities. Thus it is possible to find an approximate analytic expression for the continuum resonance damping $\gamma = \Im \left ( \delta \omega \right )$. This expression can then be applied to \TAEs\ calculated using a shooting method, which ignores the discontinuity in these modes due to resonances. The perturbation technique described assumes that the first order correction to the \TAE\ eigenfunction is small, implying small damping. Terms involving the product of two or more perturbed quantities are therefore neglected in the perturbative calculation.


Such an analytical approach to calculating continuum resonance damping could provide greater insight into its dependence on equilibrium parameters than the purely numerical methods described below. Borba \textit{et al.}\ compute continuum resonance damping using a perturbative technique applied in the code NOVA-K in a study comparing the overall damping calculated by various \MHD\ codes and experiment \cite{influence_of_plasma_shaping_on_damping}. This total damping had significant contributions from both continuum and radiative damping components. It was found that the total damping computed by \NOVA , which included continuum damping, showed reasonable agreement with other codes and experiment. However, to our knowledge, the results of the perturbative method detailed in reference \cite{continuum_damping_of_low_n_TAEs} have not previously been directly compared with those of accepted numerical techniques. Such a comparison is carried out in section ~\ref{sec:perturbative_vs_resistive}.

The ideal \MHD\ normal mode code \NOVA\ has previously been adapted to use this perturbative approach to calculate continuum resonance damping \cite{double_gap_AEs}. The code was used to calculate the very small damping due to the continuum resonance of a double-gap Alfv\'{e}n eigenmode (\DGAE). Subsequent authors have applied this code to calculate continuum damping for other Alfv\'{e}n eigenmodes \cite{interpretation_of_finite_beta_effects_in_RSAE_theory,comparrison_of_measured_and_calculated_TAE_damping_rates}.

\subsection{Resistive method}
In resistive \MHD\ the continuum resonance damping corresponds to a component of resistive damping which is independent of the resistivity. In the limit of resistive \MHD\ the \cite{computational_approach_to_continuum_damping_in_3D_published}:
\begin{equation}
-i \delta \nabla_\perp^2 \left ( \frac{\omega^2}{v_A^2} \nabla_\perp^2 \left ( E_m r \right ) \right )
\end{equation}
where the operator $\nabla_\perp^2$ is approximately $\frac{d^2}{dr^2} + \frac{1}{r} \frac{d}{dr} - \frac{m^2}{r^2}$. Inclusion of this term in the \TAE\ wave equation results in complex eigenfrequencies, with the imaginary component representing resistive damping. Continuum resonance damping is found by determining a limiting value which resistive damping approaches as the resistivity parameter is reduced. Unfortunately, the addition of the resistive term increases the order of the differential equation that must be solved. Hence, applying this resistive technique is relatively computationally intensive. 

Use of the resistive method requires that resistivity be sufficiently large that at least two discretised continuum modes are excited \cite{global_waves_in_cold_plasmas} and that the resistive contribution to the damping is negligible except near the resonance. These competing requirements potentially limit the ability to determine very small values of continuum damping using this method. While the former condition can be addressed by sufficiently increasing the grid resolution, this will increase computational requirements and potentially make the calculation impractical. By contrast, the perturbative technique is expected to be valid only where the continuum damping is small. Thus, this method potentially complements the use of resistive numerical codes for calculating continuum resonance damping.

\subsection{Complex contour method}
Continuum resonance damping can also be determined by solving the \TAE\ wave equation from ideal \MHD\ over an appropriate complex contour \cite{computational_approach_to_continuum_damping_in_3D_published}. As in the resistive technique, complex eigenfrequencies are found wherein the imaginary part represents the damping. An integration path is chosen in order to circumvent the poles due to continuum resonances. For real $\omega$ these poles would be found on the real axis, as before. However, when the \TAE\ frequency $\omega$ acquires an imaginary component, the poles become complex. The causality condition specifies which side of the poles the contour must lie on, similar to the case of Landau damping \cite{landau_damping}. This condition implies that all poles due to resonances must be between the complex contour and the real axis. Other poles, which are not physically meaningful, may result from the analytic continuation of equilibrium quantities. The contour must also remain on the same side of these as the real axis to avoid spurious contributions to the eigenmode and eigenfrequency. Providing that these conditions are met, $\omega$ will be independent of the contour used. The contour technique has been shown to agree very closely with the accepted method of performing damping calculations using resistive codes \cite{computational_approach_to_continuum_damping_in_3D_published}.

\section{Comparison of perturbative and contour techniques} \label{sec:perturbative_vs_contour}

In this section we compare the level of damping predicted by the perturbative method described by Berk et al. with that predicted by the contour method described by K\"{o}nies and Kleiber. Both methods were applied to calculate the damping of a \TAE\ due to two-mode coupling in a large-aspect-ratio tokamak with circular cross-section. Coupling of the $\left ( n,m \right ) = \left ( 2,2 \right )$ and $\left ( 2,3 \right )$ harmonics is considered, where $m$ is the poloidal mode number. For this case, the \TAE\ wave equation can be approximated using equation (\ref{eq:wave_equation}), which is used by both techniques to calculate the \TAE.

\begin{figure}[h] 
\centering 
\includegraphics[width=80mm]{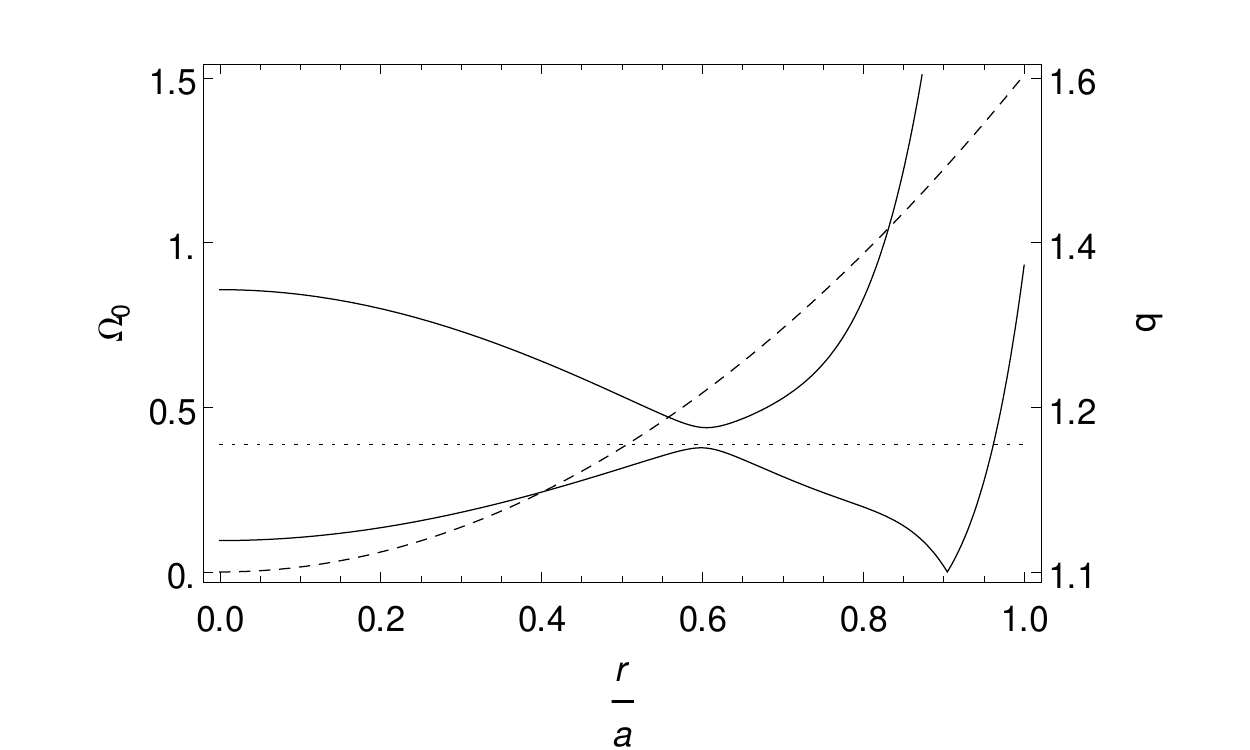} 
\caption{\label{fig:continuum} 
Continuum resonance frequency as a function of radial position $\frac{r}{a}$ (solid line, left scale), considering only the $\left ( n,m \right ) = \left ( 2,2 \right )$ and $\left ( 2,3 \right )$ harmonics. The frequency of the \TAE\ due to the interaction of these harmonics is indicated (dotted line) as is the $q$ profile (dashed line, right scale). The equilibrium parameters in this case are: $q_0=1.05$, $q_a=1.6$, $\epsilon=0.1$, $\Delta _1=0.8$ and $\Delta _2=0.1$.
} 
\end{figure}

The coupled differential equations were solved using a shooting method similar to the one outlined in \cite{continuum_damping_of_low_n_TAEs}, imposing the condition that $E_m$ be finite at the magnetic axis, $r=0$, and zero at the plasma edge,$r=a$, (corresponding to a perfectly conducting wall). We use density and safety-factor profiles similar to those used by K\"{o}nies and Kleiber \cite{computational_approach_to_continuum_damping_in_3D_published} for the large-aspect-ratio circular cross-section tokamak
\begin{equation}
q\left (r \right )=q_0 + \left ( q_a - q_0 \right )  \left (\frac{r}{a} \right )^2
\end{equation}
\begin{equation}
n\left (r \right )=\frac{n_0}{2} \left (1-\tanh \left (\frac{\frac{r}{a}-\Delta_1}{\Delta_2}\right ) \right )
\end{equation}
where the subscripts $0$ and $a$ refer to the magnetic axis and plasma edge respectively. The parameters $\Delta_1$ and $\Delta_2$ are respectively the normalised radial location of the density jump and scale length of the density gradient.  This choice of profiles results in a \TAE\ mode which has a continuum resonance at some location between the avoided crossing and the plasma edge. The continuum spectrum for this two-mode coupling case and these pressure and density profiles is plotted in figure \ref{fig:continuum}.

\begin{figure}[h] 
\centering 
\includegraphics[width=80mm]{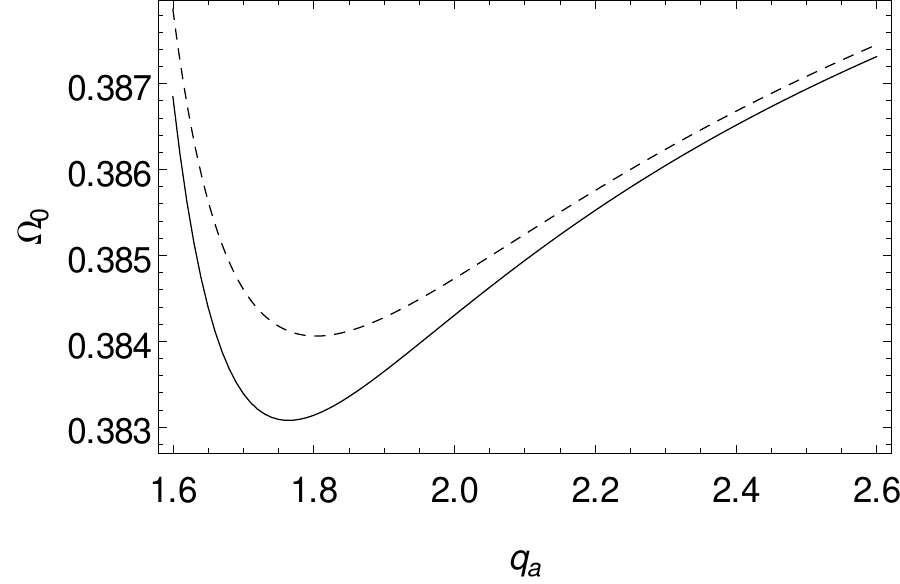} 
\includegraphics[width=80mm]{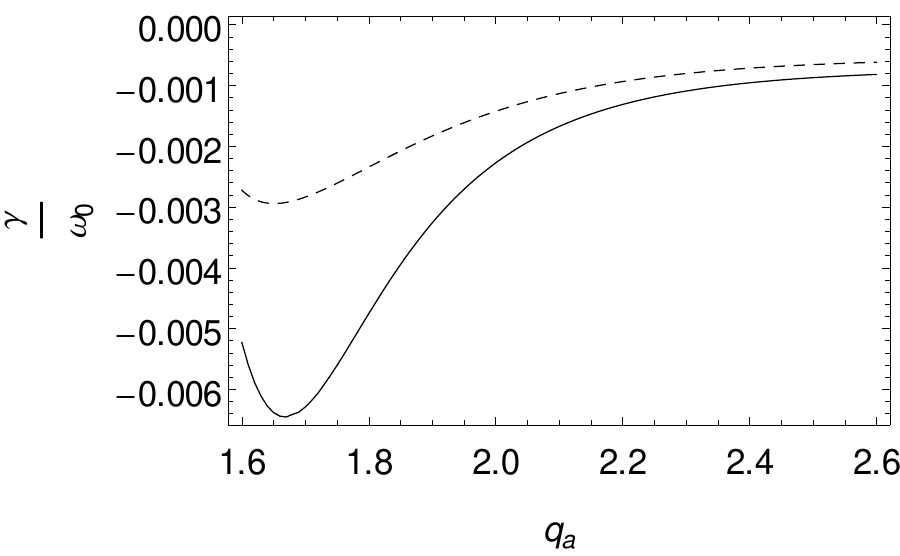} 
\caption{\label{fig:comparevsqa} 
Eigenfrequency and damping for varying $q_a$ calculated using the perturbative method (solid line) and complex contour method (dashed line). The other equilibrium parameters are fixed:  $\epsilon=0.1$, $\Delta _1=0.8$ and $\Delta _2=0.1$. $\Omega$ is plotted in figure (a) and $\frac{\gamma}{\omega_0}$ is plotted in figure (b). 
} 
\end{figure}

\begin{figure}[h] 
\centering 
\includegraphics[width=80mm]{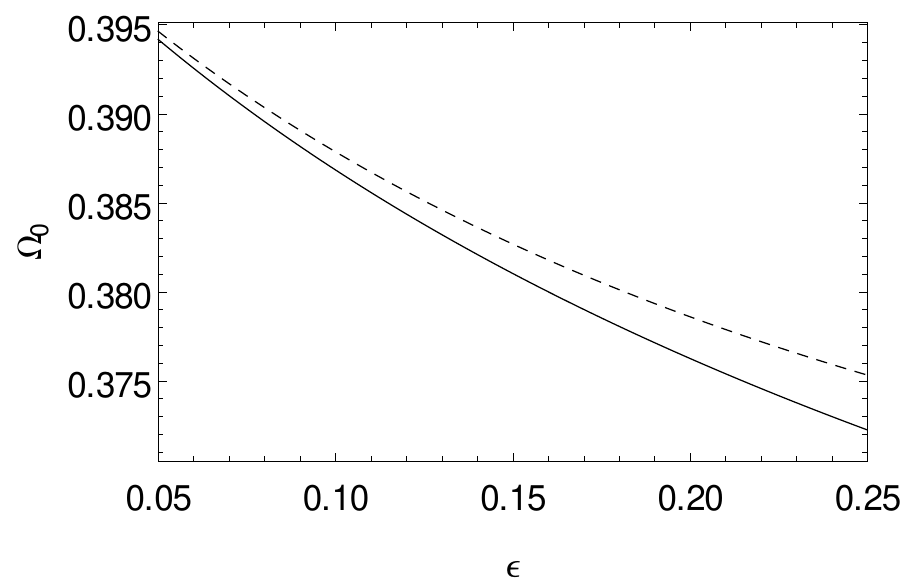} 
\includegraphics[width=80mm]{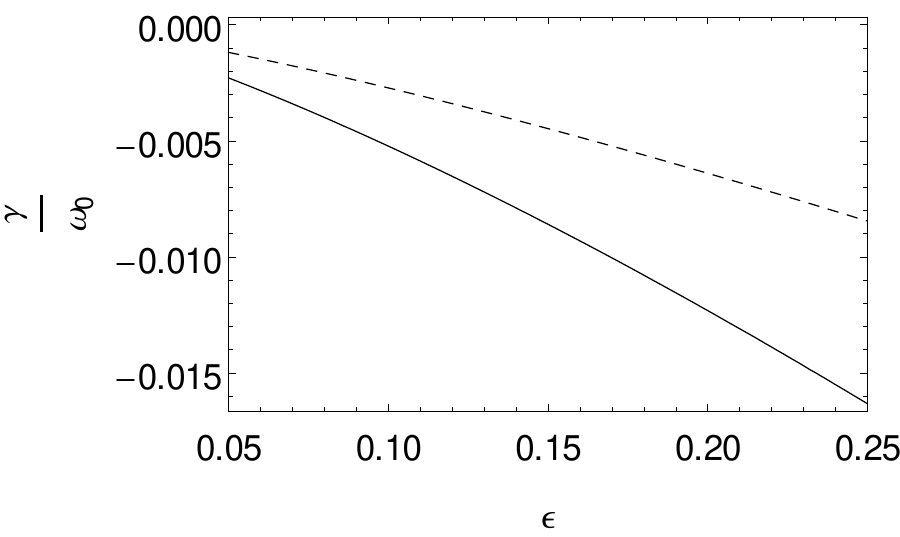}
\caption{\label{fig:comparevsepsilon} 
Eigenfrequency and damping for varying $\epsilon$ calculated using the perturbative method (solid line) and complex contour method (dashed line). The other equilibrium parameters are fixed: $q_a=1.6$, $\Delta _1=0.8$ and $\Delta _2=0.1$. $\Omega$ is plotted in figure (a) and $\frac{\gamma}{\omega_0}$ is plotted in figure (b). 
} 
\end{figure}

\begin{figure}[h] 
\centering 
\includegraphics[width=80mm]{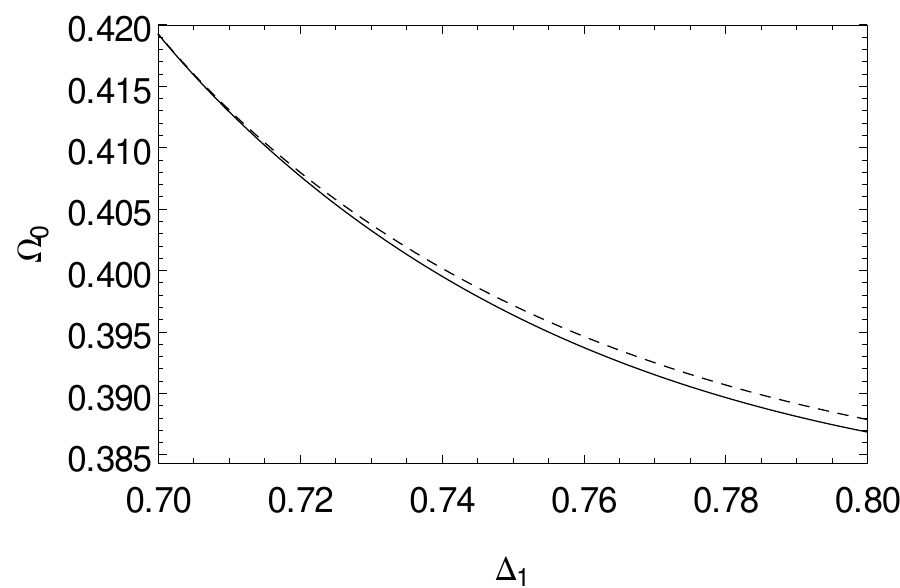} 
\includegraphics[width=80mm]{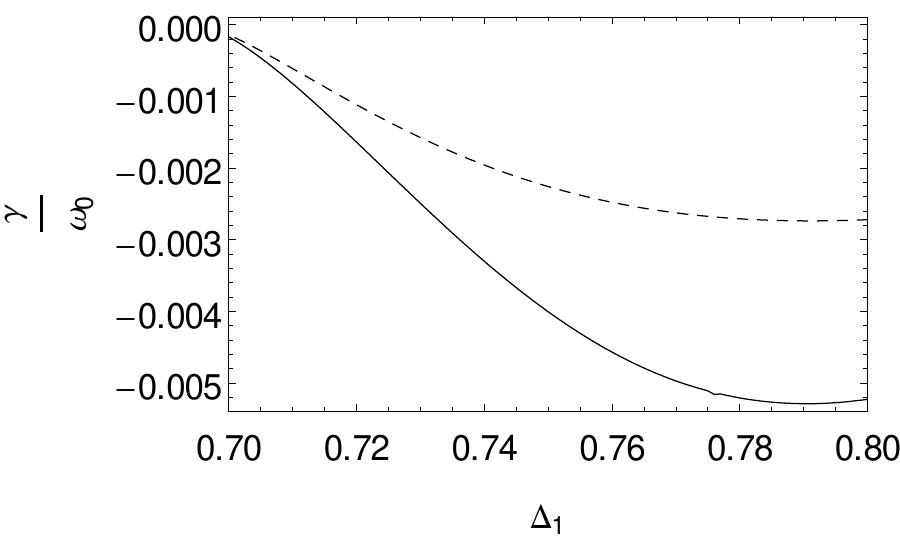}
\caption{\label{fig:comparevsDelta1} 
Eigenfrequency and damping for varying $\Delta_1$ calculated using the perturbative method (solid line) and complex contour method (dashed line). The other equilibrium parameters are fixed: $q_a=1.6$, $\epsilon=0.1$ and $\Delta _2=0.1$. $\Omega$ is plotted in figure (a) and $\frac{\gamma}{\omega_0}$ is plotted in figure (b). 
} 
\end{figure}

\begin{figure}[h] 
\centering 
\includegraphics[width=80mm]{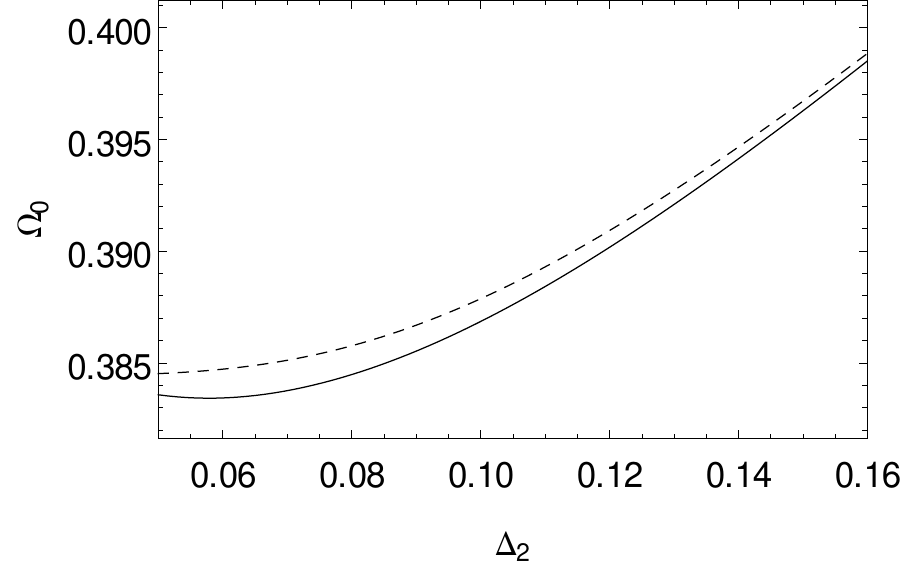} 
\includegraphics[width=80mm]{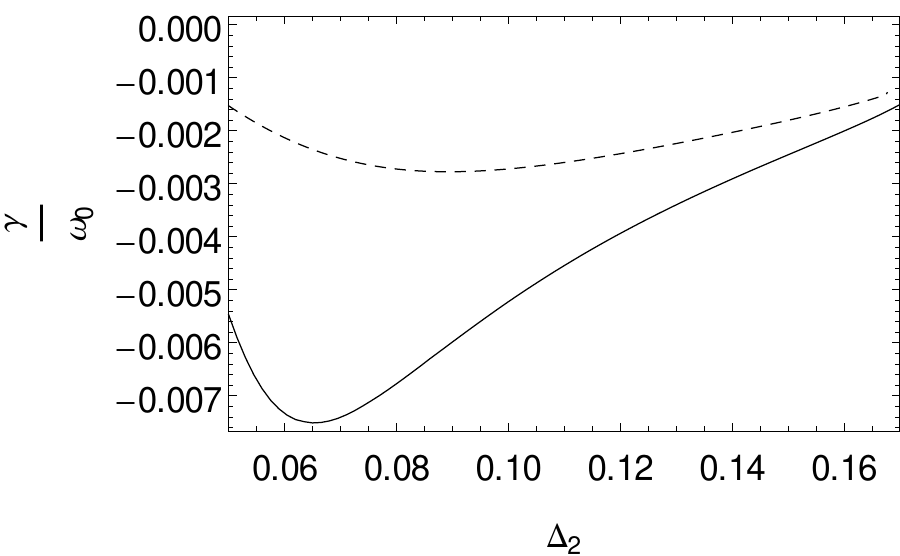}
\caption{\label{fig:comparevsDelta2} 
Eigenfrequency and damping for varying $\Delta_2$ calculated using the perturbative method (solid line) and complex contour method (dashed line). The other equilibrium parameters are fixed: $q_a=1.6$, $\epsilon=0.1$ and $\Delta _1=0.8$. $\Omega$ is plotted in figure (a) and $\frac{\gamma}{\omega_0}$ is plotted in figure (b). 
} 
\end{figure}

Figures \ref{fig:comparevsqa} to \ref{fig:comparevsDelta2} compare the results of the perturbative and contour methods for varying $q_a$ (figure \ref{fig:comparevsqa}), $\epsilon$ (figure \ref{fig:comparevsepsilon}), $\Delta _1$ (figure \ref{fig:comparevsDelta1}) and $\Delta _2$ (figure \ref{fig:comparevsDelta2}) while all other equilibrium parameters are held constant. For each case examined, the normalised \TAE\ eigenfrequency $\Omega_0 = \frac{\omega_0 R_0}{v_A\left (r=0 \right )}$ and damping ratio $\frac{\gamma}{\omega_0}$ were computed, where $\omega_0$ and $\gamma$ respectively represent the real and imaginary components of $\omega$. The safety factor at the magnetic axis is chosen to be $q_0 = 1.05$. Varying $q_a$ alters the $q$-profile, which in turn alters the location of the resonance and that of the continuum gap which produces the mode. The mode tends to be peaked in the vicinity of the gap, so that a greater separation between the gap and continuum tends to correspond to lower amplitudes at the resonance and reduced damping. However, for the low $n$ case considered here, \TAE\ modes have a relatively large radial extent and cannot necessarily be thought of as being localised at the gap. The parameter $\epsilon$ determines the strength of the coupling between poloidal modes. This parameter therefore affects the size of the gap, but does not significantly affect its location or that of the resonance. By contrast, variation in $\Delta _1$ and $\Delta _2$ both alter the density of the plasma near the edge. As the density determines the Alfv\'{e}n speed, these parameters affect the location of the resonance. The parameter $\Delta_2$ determines how rapidly density decreases around $r=\Delta_1$. It therefore affects how rapidly the continuum frequency varies with the radial coordinate and hence the radial variation of continuum frequency near the resonance.

\begin{figure}[h] 
\centering 
\includegraphics[width=60mm]{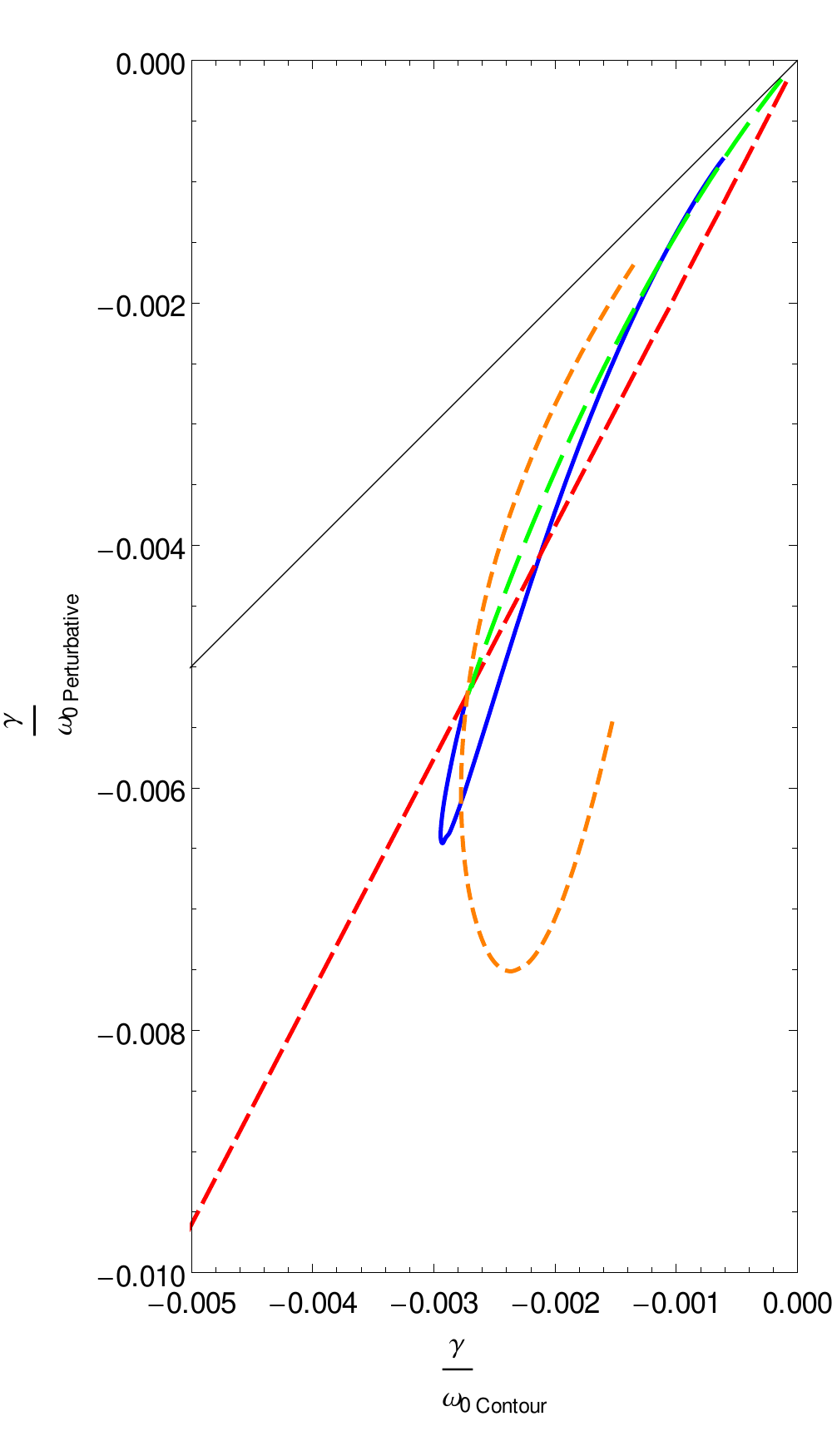} 
\includegraphics[width=60mm]{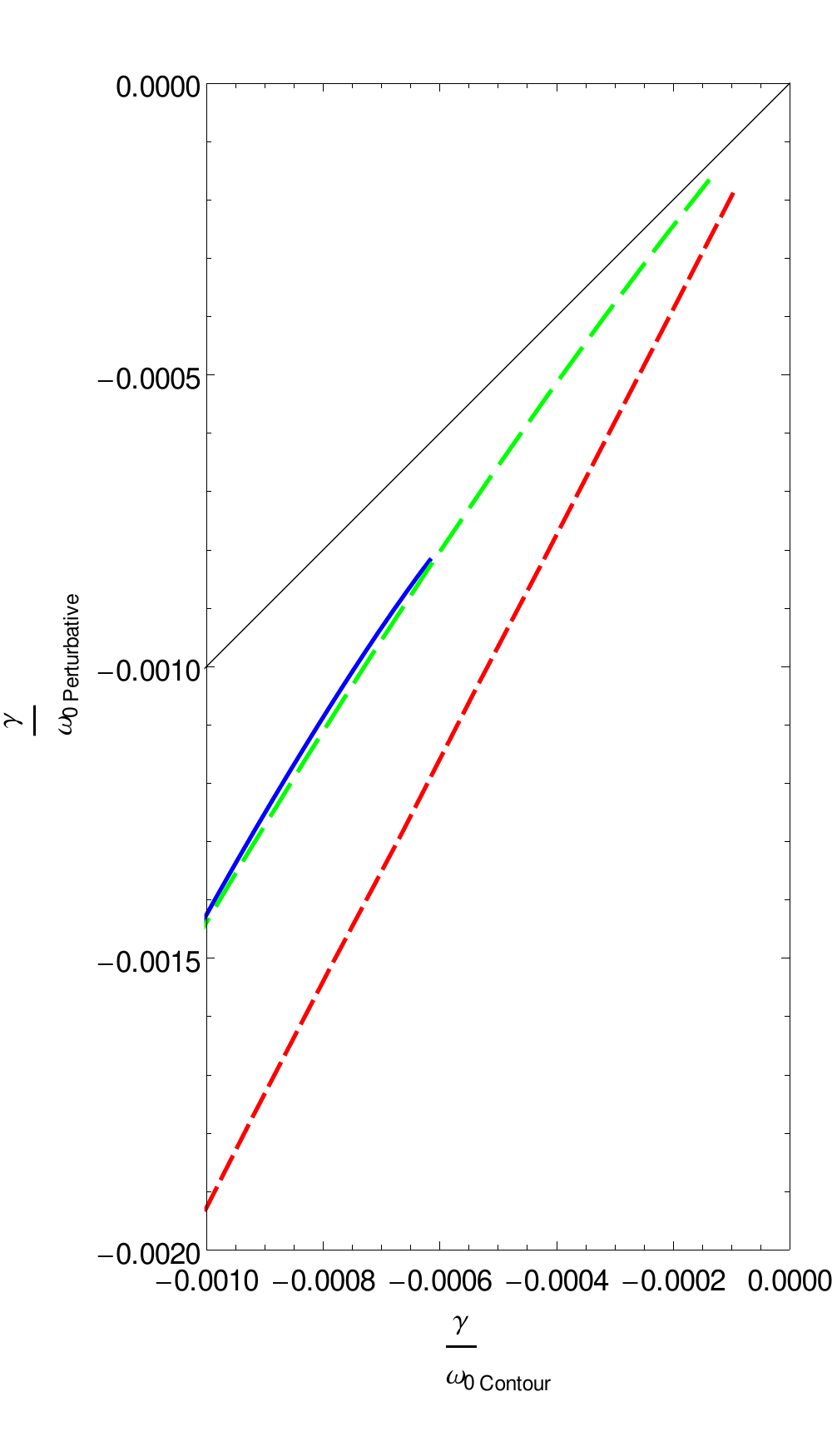} 
\caption{\label{fig:BerkvsKon} 
Damping calculated using the perturbative method against that calculated using the complex contour method for varying $q_a$ (blue, solid line), $\epsilon$ (red, medium dashed line), $\Delta_1$ (green, long dashed line) and $\Delta_2$ (orange, short dashed line). The values of equilibrium quantities are those stated in figure \ref{fig:comparevsqa}, figure \ref{fig:comparevsepsilon}, figure \ref{fig:comparevsDelta1} and figure \ref{fig:comparevsDelta2} respectively. A thin solid black line indicates where the two quantities are equal. Detail from figure (a) is shown in figure (b).
} 
\end{figure}

The contour used in the complex contour method is parameterised as $ \frac{r}{a}=x+i \alpha x \left ( 1-x \right ) $, where $x \in \left (0,1 \right )$. The parameter $\alpha$ is adjusted between damping calculations so that the imaginary deformation of the contour is proportional to the damping rate. This ensures that the contour remains on the side of the resonance pole of equation (\ref{eq:wave_equation}) specified by the causality condition and that spurious poles are avoided.

In figure \ref{fig:comparevsqa} to figure \ref{fig:comparevsDelta2} the estimate of damping provided by the perturbative method is seen to be consistently larger than the value calculated by the contour method. This discrepancy is most significant where the contour technique calculates a large value for the damping ratio. This corresponds to the cases where the \TAE\ mode is localised near the resonance and the complex discontinuity in the wave function across the resonance is large. Hence, it appears that the discrepancy may result from a significant perturbation to the flux function $C_m$ at the resonance, as suggested by Berk \textit{et al.} \cite{continuum_damping_of_low_n_TAEs}. This function depends on both $\frac{dE_m}{dr}$ and $\omega$, which change when the effect of the resonance is included.

The damping ratios calculated using the two techniques are compared directly in figure \ref{fig:BerkvsKon}. Surprisingly, the damping ratio estimates that are obtained using the perturbative technique do not appear to approach those of the complex contour method as damping becomes very small. Instead, the ratio between the two damping estimates takes an approximately constant value, which depends on the equilibrium parameter being varied. This indicates that the relative change in $C_m$ at the resonance due to inclusion of the effect of the discontinuity remains significant even for very small damping.

Nevertheless, there is qualitative agreement between the two techniques regarding how damping  varies with $q_a$, $\epsilon$ and $\Delta_1$. The perturbative and contour methods approximately agree on the values of $q_a$ and $\Delta_1$ for which damping is maximised and minimised over the ranges of values examined. The two methods are both able to account for changes in damping as the position of the gap giving rise to the \TAE\ mode and that of the resonance are changed. Similarly, both techniques show that the damping rate increases approximately linearly with $\epsilon$. This is to be expected, as increasing $\epsilon$ corresponds to increasing coupling between modes due to toroidicity. In each of these three parameter scans, the values of damping calculated using the two methods are approximately linearly related, as illustrated in figure \ref{fig:BerkvsKon}. This is not true when $\Delta_2$ is varied. For this parameter scan, the damping ratios cannot be overlaid by linearly rescaling the calculated damping ratio. This indicates that the higher order terms neglected by the perturbative approach are not linearly related to the first order term which is considered. Unlike the other parameter scans, the perturbative technique does not agree closely with the complex contour technique regarding which value of $\Delta_2$ maximises damping.

To quantify the degree of linearity in their relationship, correlation coefficients are computed for the damping ratios determined using the two techniques over the ranges of parameters indicated in figures \ref{fig:comparevsqa}\ to \ref{fig:comparevsDelta2}. It is found that the correlation coefficients for these estimates as $q_a$, $\epsilon$ or $\Delta_2$ are each greater than $0.99$, demonstrating a high degree of linear correlation. By contrast, the correlation coefficient for the estimates as $\Delta_2$ is varied is $0.59$. Figure \ref{fig:BerkvsKon} illustrates the approximately linear relationship in between damping estimates for variation in $q_a$, $\epsilon$ and $\Delta_1$ and the non-linear relationship for variation in $\Delta_2$.

The two techniques also calculate different results for the real part of the frequency of the \TAE\ mode. The real component calculated using the perturbative approach is found to be smaller than that calculated using the complex contour method for all of the sets of parameters examined. This difference is of the same order of magnitude as the damping and scales with damping. This observation implies that the \TAE\ frequency expansion with respect to the perturbed quantities has a contribution from neglected higher order terms that has a real component comparable to the first order imaginary term considered in the formalism of Berk \textit{et al.}.Such a contribution is consistent with a significant change in $C_m$ at the continuum resonance due to inclusion of the discontinuity there.

\begin{figure}[h] 
\centering 
\includegraphics[width=80mm]{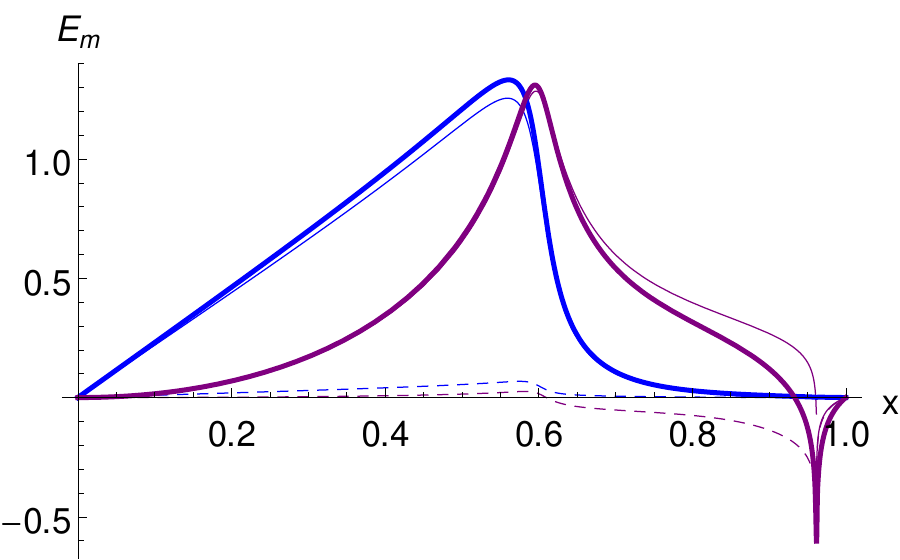} 
\includegraphics[width=80mm]{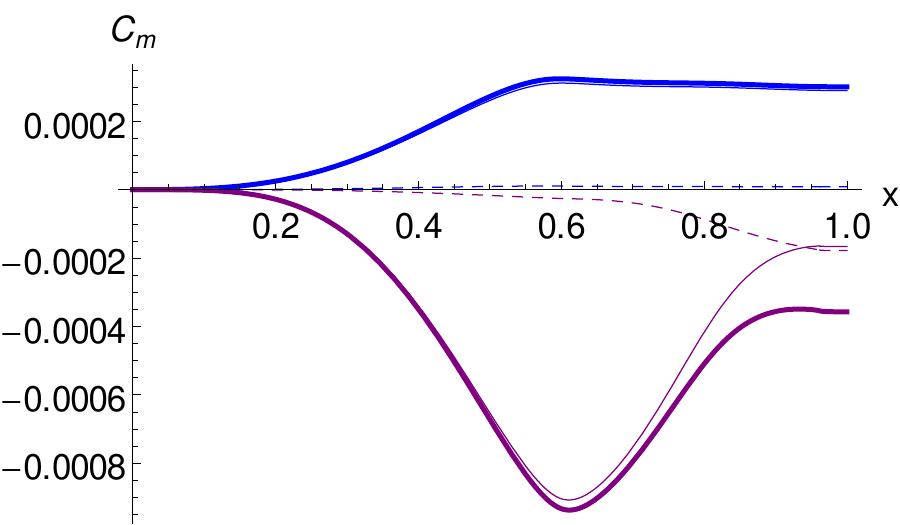} 
\caption{\label{fig:compareBerkChu} 
$E_m$ and $C_m$ calculated by equating these quantities on both sides of resonances as done by Berk et al. \cite{continuum_damping_of_low_n_TAEs} (thick line) and by applying resonance pole contributions calculated analytically, similar to Chu et al. \cite{numerical_study_of_Alfven_continuum_damping_of_AEs} (thin line, real solid, imaginary dashed). The $n=2$, $m=2$ (blue) and $n=2$, $m=3$ (purple) harmonics are shown. This case corresponds to the equilibrium parameters $q_a=1.6$, $\epsilon=0.1$, $\Delta _1=0.8$ and $\Delta _2=0.1$.
} 
\end{figure}

As the contour technique has been shown to be accurate \cite{computational_approach_to_continuum_damping_in_3D_published}, the discrepancy between this and the perturbative technique must arise from the approximations made in the latter technique. It is thought that the main source of error in using the formalism of Berk et al. is inaccuracy in determining the variable $C_m$ at the resonance \cite{continuum_damping_of_low_n_TAEs}. To verify this explanation, the first order solution for a \TAE\ using the method described above is compared with a solution found by analytically calculating the contribution of poles at resonances, similar to the method described by Chu \textit{et al.} \cite{numerical_study_of_Alfven_continuum_damping_of_AEs}. \TAE\ modes calculated using each method are compared in figure \ref{fig:compareBerkChu}. The eigenfrequency and damping calculated by computing pole contributions analytically agree closely with those calculated by the complex contour method. However, by analytically including pole contributions the variables $E_m$ and $C_m$ can be found for positions on the real axis.

In the case examined, $C_3$ has a significant imaginary component and smaller real component near the resonance location when the contribution of the corresponding pole is considered. This corresponds to changes in the real and imaginary components of $E_3$ in this region. The discontinuity in $E_m$ at the resonance has the same effect as significantly changing the boundary conditions for the problem. The profiles for $C_2$ and $E_2$ are similar in both cases, as this mode is less affected by the resonance, which is with a primarily $m=3$ branch of the continuum.

\section{Comparison of perturbative and resistive methods} \label{sec:perturbative_vs_resistive}
The damping calculated using a perturbative method is compared to that calculated using a resistive method for a more detailed tokamak model. The large aspect ratio approximation is not made in this case, so that the effect of terms in the \TAE\ mode equation which are of higher order with respect to aspect ratio is included. The effect of including more harmonics, in addition to the dominant \TAE\ components considered previously, is also incorporated.

Perturbative and resistive calculations were performed using the finite element \MHD\ codes \NOVA\ \cite{nova_a_nonvariational_code,double_gap_AEs} and \CKA\ \cite{computational_approach_to_continuum_damping_in_3D_published}, respectively. In these codes, the problem is discretised using the Galerkin method. A weak formulation is obtained by multiplying the relevant force operator equation by a test function and integrating over the radial coordinate. This leads to an equation in terms of a bilinear form, which can be discretised by expressing the test function and solution function as linear sums of a set of basis functions. Thus it is possible to approximate the problem as a generalised matrix eigenvalue equation, which can be solved computationally to find the \TAE\ frequency and wave function.

\begin{figure}[h] 
\centering 
\includegraphics[width=80mm]{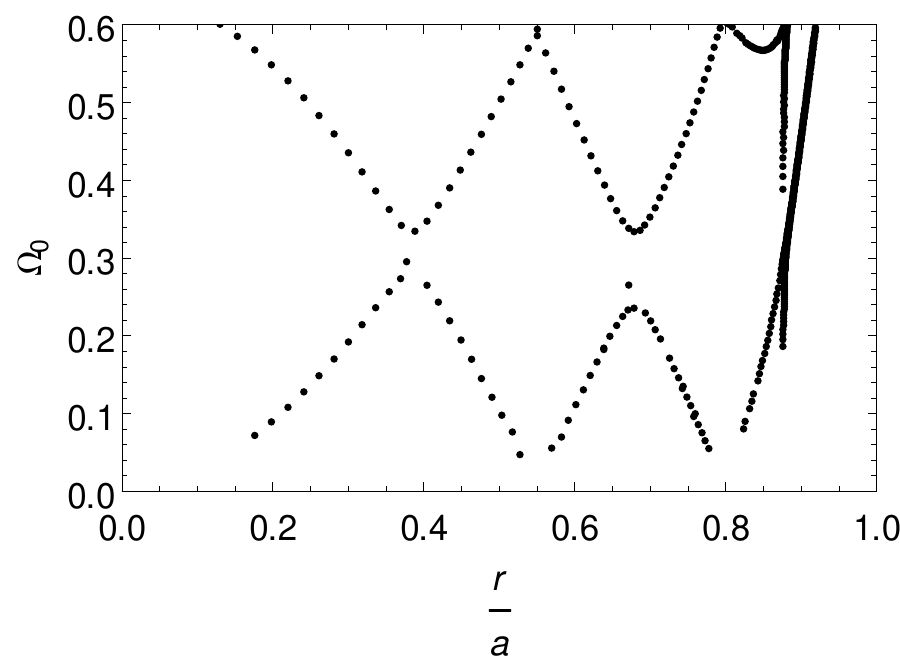} 
\caption{\label{fig:CKA_continuum} 
Shear Alfv\'{e}n spectrum for $n=6$ with $q$ and $n$ as described in section \ref{sec:perturbative_vs_resistive}. A \TAE\ exists at $\Omega_0 \approx 0.295$ in the gap due to the avoided crossing of the $\left (n,m \right ) = \left ( 6,9 \right )$ and $\left ( 6,10 \right )$ branches. This spectrum was calculated using \CKA\ with $20$ poloidal grid points and $800$ radial grid points, with clustering at $r=0.876$ near the location of the \TAE\ continuum resonance.
}
\end{figure}

The safety factor used for the comparison has a similar form to that used in the previous section, $q\left (s \right ) = q_0 + \left ( q_a - q_0 \right ) s$, where $s\approx \left ( \frac{r}{a} \right )^2$ is the normalised flux. In this case the parameters $q_0=1.5$ and $q_a=2.0$ are used. The density profile has the form $n\left (s \right ) = n_0 \left (1-s^{\beta} \right )^{\alpha}$, where the values of the parameters are $\alpha = 7.0$ and $\beta = 7.5$. The inverse aspect ratio has been chosen to be $\epsilon = 0.2$. These parameters and functions have been chosen differently to those in the previous section, in an attempt to ensure the best possible convergence of the damping rate with respect to radial grid resolution. Thus the parameters are selected in order that the \TAE\ is localised away from the resonance, giving a small continuum damping value for which the perturbative approach is most likely to be valid. The two codes were used to compute a \TAE\ within the gap between the $\left (n,m \right ) = \left ( 6,9 \right )$ and $\left ( 6,10 \right )$ branches of the shear Alfv\'{e}n continuum. The shear Alfv\'{e}n spectrum for this case, calculated using \CKA, is shown in figure \ref{fig:CKA_continuum}.

The eigenvalue of the \TAE\ in the resistive calculation is $\omega ^2 = (0.0871345753 - i 0.0000000136) \left ( \frac{v_A}{R_0}\right )^2$, representing a normalised real eigenfrequency component of $\Omega_0 = 0.295185662$ and a damping ratio of $\frac{\gamma}{\omega_0}=-7.81 \times 10^{-7}$. This value is found to have converged for $800$ radial grid points, $20$ poloidal grid points and an artificial damping parameter of $\delta = 10^{-12}$ (as described in \cite{computational_approach_to_continuum_damping_in_3D_published}). This convergence is demonstrated by a change in calculated damping ratio of $0.30\%$ when the radial grid resolution is reduced to $600$ points, a change of $1.7\%$ when the poloidal grid resolution is reduced to $15$ points, and a change of $0.15\%$ in this ratio when the damping parameter is increased to $10^{-11}$. Radial grid points are clustered near the location of the resonance in these calculations, in order to resolve the rapid variation of the wave function in this region.

\begin{figure}[h] 
\centering 
\includegraphics[width=80mm]{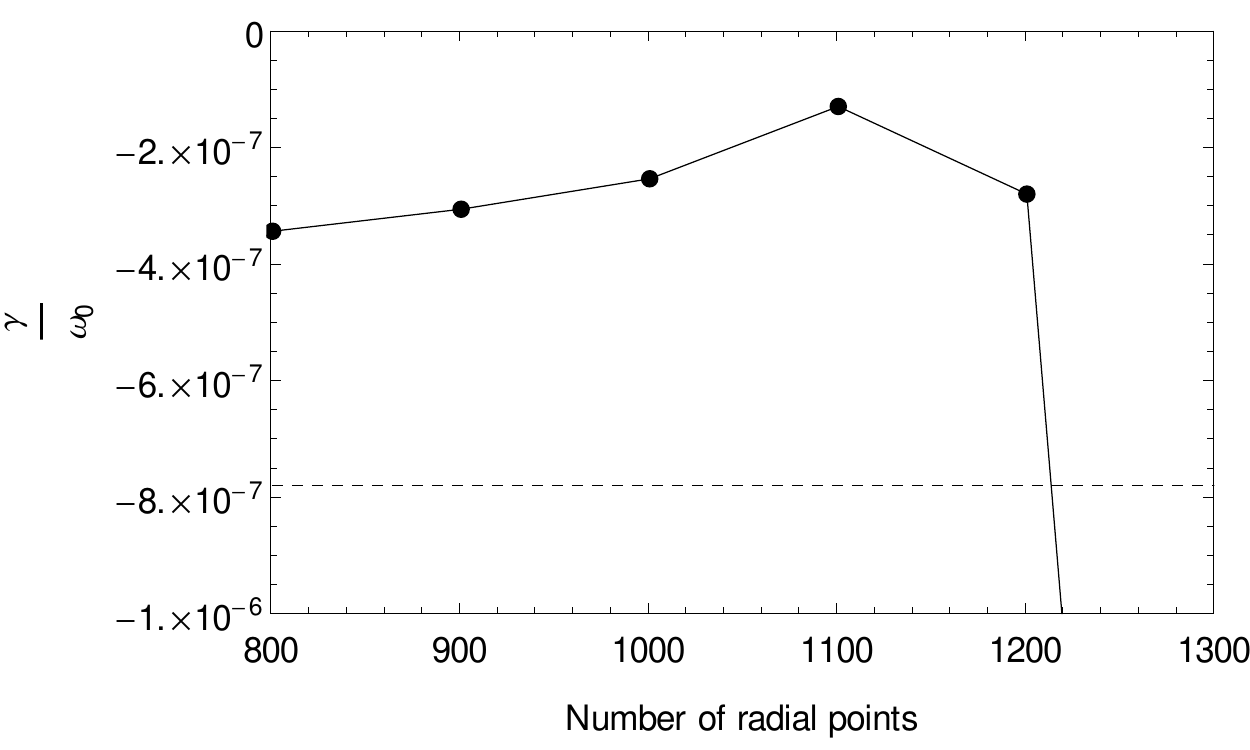} 
\includegraphics[width=80mm]{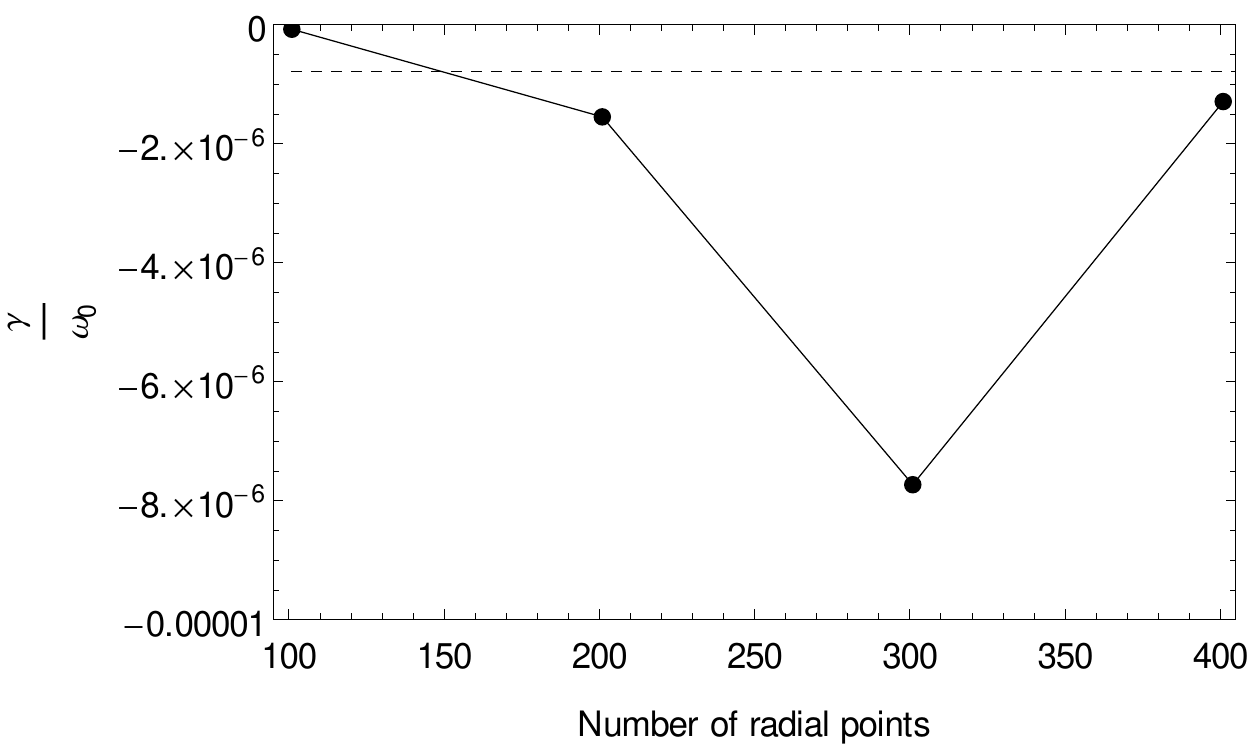} 
\caption{\label{fig:NOVA_convergence} 
Convergence study of continuum resonance damping calculated by \NOVA\. The damping ratio calculated for various radial grid resolutions (solid line with circles indicating data points) is shown along with the converged value from the resistive calculation performed by \CKA (dashed line).
} 
\end{figure}

By contrast, the estimate of continuum damping obtained using the perturbative method does not converge satisfactorily, as shown in figure \ref{fig:NOVA_convergence}. For radial grid resolution of between $101$ and $401$ radial grid points, very large variation in the computed damping ratio is observed. Between $801$ and $1201$ radial grid points there is lesser, but still significant, variation in the continuum resonance damping computed using \NOVA. This suggests a damping ratio of between $1 \times 10^{-7}$ and $4 \times 10^{-7}$, significantly smaller than the resistive estimate. However, further increasing the resolution to $1301$ and then $1401$ grid points results in a rapid increase in the calculated damping ratio. In comparison, the real part of the \TAE\ frequency computed by \NOVA\ converges to $\omega_0 = 0.29572\left ( \frac{v_A}{R_0}\right )$ for $101$ radial grid points, agreeing closely with the resistive calculation using \CKA\ (to within $0.2\%$). The mode structure also converges and is in close agreement with \CKA\, except in the region near the resonance. A sharp peak is observed at the resonance, with its size generally increasing with estimated damping. The wave functions are not symmetrical about the singularity in the region surrounding the peak, contrary to the logarithmic form of the singularity indicated by a Frobenius expansion. This results in an erroneus discontinuity in the wave function at the singularity.

The convergence of the perturbative continuum damping estimate with respect to radial grid resolution has been investigated for several additional cases. Using \NOVA\, \TAEs\ were found for various different values of $n$ and different $q$ and $\rho$ profiles. The effect of clustering radial grid points near the resonance was also examined. In each case the continuum damping estimate did not satisfactorily converge with radial grid resolution. Indeed, it was found that addition of a single additional grid point could sometimes dramatically change the continuum damping calculated.

The failure of the continuum damping estimate of \NOVA\ to converge appears to be due to the failure of the ideal \MHD\ code to properly incorporate the singularity due to the resonance. \NOVA\ represents the wave function using a linear combination of B-spline functions. Splines are piecewise polynomials of order $n$ which are differentiable to order $n-1$ at interior points. B-splines have the distinction of having minimal support\cite{applications_of_b_splines}, simplifying the matrix eigenvalue problem by minimising the number of non-zero elements encountered. Polynomial functions cannot accurately represent the logarithmic singularity of the \TAE\ at the resonance. In an abstract sense, the \TAE\ solution found using a finite element is a projection of the exact solution onto the solution space spanned by this basis set. When the bilinear form operates on the difference between the exact and finite element solutions and any element of the basis set, the result is zero, a condition referred to as Galerkin orthogonality. Therefore, the location of the radial grid points around the singularity, which determines the basis functions, significantly affects the \TAE\ solution obtained in this region. This means that it is not possible to accurately or consistently compute the damping using the perturbative approach, which relies on the value of the \TAE\ flux function at the resonance. This problem would be encountered by any finite element implementation of the perturbative method which does not use appropriate singular finite elements to represent the function near resonances.

As the radial grid resolution is increased, the logarithmic singularity is not resolved. The piecewise polynomial basis functions cannot accurately represent the singular wave function in a region encompassing the grid points nearest to the resonance. This leads to a grid-dependent jump in the wave function across the resonance. This jump does not decay with increasing grid resolution, despite the reduction in the region where the logarithmic behaviour is poorly represented. This is because dilation of a logarithmic function, $\ln \left (x \right )$, with respect to the variable $x$ merely corresponds to a translation in the function. A potential solution to the convergence issue would be to use a singular finite element to represent the singularity at this point. Another method would be to accurately determine the location of the resonance and ensure that nearby grid points are spaced symmetrically around it. Each of these methods would require foreknowledge of the location of the continuum resonance, which is itself dependent upon the \TAE\ frequency. While such measures may improve convergence, they do not address the inaccuracy inherent in the perturbative technique identified in section \ref{sec:perturbative_vs_contour}. Moreover, such modifications are beyond the scope of the present study.

Although a large number of harmonics are included in these calculations \footnote{\CKA\ uses splines to represent the wave function in all dimensions, including poloidally. Thus, the number of poloidal harmonics that \CKA\ can effectively represent is determined by the poloidal grid resolution chosen.}, it was found that the dominant components of the \TAE\ were the $\left ( n,m \right ) = \left ( 6,9 \right )$ and $\left ( 6,10 \right )$ harmonics, as expected. Therefore, it seems unlikely that considering additional modes significantly alters the damping calculated. This was supported by the observation that adding additional modes to the simplified perturbative \TAE\ calculation of the previous section does not significantly alter the frequency, damping or structure of the mode.

\section{Conclusion}
The perturbative technique examined can provide a qualitative understanding of how continuum resonance damping varies with equilibrium parameters. However, the magnitude of damping calculated by the perturbative technique generally has a significant error with respect to the accepted value. This is found even for cases with simplified geometry and very small damping. The observed discrepancy results from violation of the assumption in the perturbative method that the flux function $C_m$ does not change significantly due to the inclusion of the discontinuity at the resonance. Moreover, the perturbative method has inherently poor convergence when applied to ideal \MHD\ finite element codes without appropriate singular basis functions. The estimate obtained depends strongly upon the wave function solution near the resonance, which does not converge due to these codes' failure to accurately model the singularity due to the resonance. It is possible to accurately calculate continuum damping in ideal \MHD\ using a method which includes the imaginary part of this singularity, such as the complex contour method or analytically calculating contributions from complex poles.

\section*{Acknowledgments}

The authors gratefully acknowledge support of the Australian Research Council, through Grants No. FT0991899 and No. DP110102881 and support from the German Academic Exchange Service (DAAD) Project No. 50153864. We would also like to acknowledge the technical assistance provided by Dr. B. Seiwald.

\end{document}